\documentclass[sigconf]{acmart}
\AtBeginDocument{%
  }

\setcopyright{acmlicensed}
\copyrightyear{2018}
\acmYear{2018}
\acmDOI{XXXXXXX.XXXXXXX}
\acmConference[MM'25]{Make sure to enter the correct
  conference title from your rights confirmation email}{October 27–31,
  2025}{Dublin, Ireland}
\acmISBN{978-1-4503-XXXX-X/2018/06}

\usepackage{enumitem}
\usepackage{amsmath}
\usepackage{multirow}
\usepackage{fancyvrb}
\usepackage{graphicx}
\usepackage{subcaption}
\newcommand{\lz}[1]{{\color{red}#1}}

\begin{document}

\title{ALLM4ADD: Unlocking the Capabilities of Audio Large Language Models for Audio Deepfake Detection}

\author{Hao Gu}
\affiliation{%
  \institution{Institute of Automation, Chinese Academy of Sciences}
  \city{Beijing}
  \country{China}}
\email{guhao2022@ia.ac.cn}

\author{Jiangyan Yi\textsuperscript{\textsuperscript{$\dagger$}}}
\affiliation{%
  \institution{Department of Automation, Tsinghua University}
  \city{Beijing}
  \country{China}}
\email{yijy@tsinghua.edu.cn}

\author{Chenglong Wang}
\affiliation{%
  \institution{Taizhou University}
  \city{Taizhou}
  \country{China}}
\email{wcl519@tzc.edu.cn}

\author{Jianhua Tao}
\affiliation{%
  \institution{Department of Automation, Tsinghua University}
  \city{Beijing}
  \country{China}}
\email{jhtao@tsinghua.edu.cn}

\author{Zheng Lian}
\affiliation{%
  \institution{Institute of Automation, Chinese Academy of Sciences}
  \city{Beijing}
  \country{China}}
\email{lianzheng2016@ia.ac.cn}

\author{Jiayi He}
\affiliation{%
  \institution{Institute of Automation, Chinese Academy of Sciences}
  \city{Beijing}
  \country{China}}
\email{jiayi.he@ia.ac.cn}

\author{Yong Ren}
\affiliation{%
  \institution{Institute of Automation, Chinese Academy of Sciences}
  \city{Beijing}
  \country{China}}
\email{renyong2020@ia.ac.cn}

\author{Yujie Chen}
\affiliation{%
  \institution{Anhui University}
  \city{HeFei}
  \country{China}}
\email{e22201148@stu.ahu.edu.cn}

\author{Zhengqi Wen}
\affiliation{%
  \institution{Beijing National Research Center for Information Science and Technology,Tsinghua University}
  \city{Beijing}
  \country{China}}
\email{zqwen@tsinghua.edu.cn}
 
\renewcommand{\shortauthors}{Hao Gu et al.}
\begin{abstract}

Audio deepfake detection (ADD) has grown increasingly important due to the rise of high-fidelity audio generative models and their potential for misuse. Given that audio large language models (ALLMs) have made significant progress in various audio processing tasks, a heuristic question arises: \textit{Can ALLMs be leveraged to solve ADD?}. In this paper, we first conduct a comprehensive zero-shot evaluation of ALLMs on ADD, revealing their ineffectiveness. To this end, we propose ALLM4ADD, an ALLM-driven framework for ADD. Specifically, we reformulate ADD task as an audio question answering problem, prompting the model with the question: ``Is this audio fake or real?''.  We then perform supervised fine-tuning to enable the ALLM to assess the authenticity of query audio. Extensive experiments are conducted to demonstrate that our ALLM-based method can achieve superior performance in fake audio detection, particularly in data-scarce scenarios. As a pioneering study, we anticipate that this work will inspire the research community to leverage ALLMs to develop more effective ADD systems.  Code is available at \url{https://github.com/ucas-hao/qwen_audio_for_add.git}

\end{abstract}




\maketitle

{\let\thefootnote\relax\footnotetext{\textsuperscript{$\dagger$}Corresponding author.}}
\section{Introduction}


Over the past few years, text-to-speech (TTS) and voice conversion (VC) technologies have advanced rapidly, enabling the generation of high-fidelity, human-like speech \citep{DBLP:conf/icassp/WangFYTWQW21,DBLP:journals/corr/abs-2008-12527,DBLP:conf/coling/GuYL0Y24,DBLP:journals/csl/WangYTDNESVKLJA20}. However, these technologies can be misused for malicious purposes, such as spreading misinformation, inciting social unrest, and undermining trust in digital media \citep{DBLP:conf/cvpr/JiaLZC0JHLWL22,DBLP:conf/dada/YiTFYWWZZZRXZGW23}. Therefore, audio deepfake detection (ADD) has become an increasingly urgent and essential task that needs to be addressed \citep{DBLP:journals/corr/abs-2308-14970}.





In recent years, numerous audio deepfake detection methods have been proposed, which can be broadly categorized into two types: conventional pipeline solutions and end-to-end models \citep{DBLP:journals/corr/abs-2308-14970,DBLP:conf/dada/YiTFYWWZZZRXZGW23,DBLP:journals/tbbis/NautschWEKVTDSY21,DBLP:conf/icassp/YiFTNMWWTBFLWZY22,DBLP:conf/icassp/Tak0TNEL21}. The conventional pipeline approach, consisting of a front-end feature extractor and a back-end classifier, has been the standard framework for decades \citep{DBLP:conf/mm/GuYWR0YC024,DBLP:conf/mm/ZhangWH24,DBLP:conf/odyssey/0037Y22}. In contrast, end-to-end models employ a unified model that optimizes both the feature extraction and classification processes by operating directly on raw audio waveforms \citep{DBLP:journals/corr/abs-2406-06086,DBLP:journals/corr/abs-2110-01200,DBLP:conf/icassp/Tak0TNEL21,DBLP:journals/corr/abs-2107-12710}.

Recently, Audio Large Language Models (ALLMs) \citep{DBLP:conf/iclr/0001LLKG24,DBLP:conf/emnlp/ZhangLZZWZQ23,DBLP:journals/corr/abs-2311-07919,DBLP:journals/corr/abs-2407-10759} have demonstrated remarkable progress across a wide range of audio processing tasks, including audio captioning \citep{DBLP:conf/icassp/TangYSC0LLMZ24,DBLP:journals/corr/abs-2502-15178,DBLP:journals/corr/abs-2409-12962} and speech recognition \citep{DBLP:journals/corr/abs-2408-09491,DBLP:conf/icassp/FathullahWLJSLG24,DBLP:journals/corr/abs-2311-07919}.  Notable models include Qwen-Audio \citep{DBLP:journals/corr/abs-2311-07919}, which integrates the Whisper encoder \citep{DBLP:conf/icml/RadfordKXBMS23} with the text Qwen LLM \citep{DBLP:journals/corr/abs-2309-16609}, enabling the latter to understand audio.  However, their performance on ADD task remains unexplored. This raises a critical question: \textit{Can ALLMs effectively address the ADD task?} 



To answer this question, we present pioneering work that leverages ALLMs for ADD task. To the best of our knowledge, this is the first paper to tackle ADD using the ALLM-based approach.  First, we conduct a comprehensive quantitative evaluation of ALLM's capabilities in ADD task. Our experimental results reveal that existing ALLMs perform poorly in zero-shot fake audio detection, primarily due to the mismatch between their pretraining objectives and the fake audio detection requirements.  To enhance their performance for fake audio detection, we further propose a novel framework called ALLM4ADD. Specifically, we reformulate ADD task as an Audio Question Answering (AQA) problem, prompting the model with the question ``Is this audio fake or real?'' and instructing it to generate the correct answer.  We then employ supervised fine-tuning (SFT) to endow the ALLMs the capability to answer ``Fake'' if the query audio is fake, and conversely, ``Real'' if it is real.  Extensive experimental results demonstrate that ALLM4ADD achieves superior performance compared to existing conventional pipeline and end-to-end models, particularly in data-scarce scenarios.  These findings collectively highlight the advantages of our ALLM-based approach for fake audio detection.   

In conclusion, our main contributions are threefold:

\begin{itemize}[leftmargin=*]
    \item We conduct the first comprehensive zero-shot evaluation of ALLMs for fake audio detection, demonstrating that current ALLM models perform poorly on ADD task.
    
    \item We propose ALLM4ADD, a novel framework which reformulates ADD task as an AQA problem, to successfully endow ALLMs with fake audio detection capability.
    
    \item Extensive experiments validate the effectiveness of our method, demonstrating superior performance compared to both conventional pipeline and end-to-end baselines.  Notably, it achieves strong performance in data-scarce scenarios.
\end{itemize}

\section{Related Work} \label{related work}

\subsection{Audio Deepfake Detection Methods}
In recent years, the field of audio deepfake detection has witnessed significant advancements, focusing on distinguishing genuine utterances from AI-generated fake ones \citep{DBLP:conf/mm/GuYWR0YC024,DBLP:journals/corr/abs-2308-14970,DBLP:conf/icassp/YiFTNMWWTBFLWZY22,DBLP:conf/dada/YiTFYWWZZZRXZGW23}. Existing studies typically follow one of two paradigms: the conventional pipeline approach, which combines a front-end feature extractor with a back-end classifier \citep{DBLP:journals/corr/abs-2308-14970,DBLP:conf/mm/ZhangWH24}, or the end-to-end approach, which directly processes raw audio waveforms \citep{DBLP:journals/corr/abs-2406-06086,DBLP:conf/icassp/LiuLWLZD23}.


\begin{table*}[htbp]
\centering
\caption{Zero-shot performance of the Qwen-audio and Qwen2-audio series on the ASVspoof2019 evaluation set across various prompt templates. We present both accuracy (\%) and modified mF1-score (\%). ACC denotes accuracy.}
\label{tab:qwen-audio-performance}
\renewcommand{\arraystretch}{0.75}
\begin{tabular}{c|cccccccccc|cc}
\toprule
\multirow{2}{*}{Model} & \multicolumn{2}{c}{prompt1} & \multicolumn{2}{c}{prompt2} & \multicolumn{2}{c}{prompt3} & \multicolumn{2}{c}{prompt4} & \multicolumn{2}{c}{prompt5} & \multicolumn{2}{|c}{Average} \\
\cmidrule(r){2-3} \cmidrule(r){4-5} \cmidrule(r){6-7} \cmidrule(r){8-9} \cmidrule(r){10-11} \cmidrule(r){12-13}
 & ACC \lz{$\uparrow$}& mF1 \lz{$\uparrow$} & ACC \lz{$\uparrow$} & mF1 \lz{$\uparrow$} & ACC \lz{$\uparrow$} & mF1 \lz{$\uparrow$} & ACC \lz{$\uparrow$} & mF1 \lz{$\uparrow$} & ACC \lz{$\uparrow$} & mF1 \lz{$\uparrow$} & ACC \lz{$\uparrow$} & mF1 \lz{$\uparrow$} \\
\midrule
Qwen\text{-}audio\text{-}base & 14.91 & 6.49 & 12.13 & 1.86 & 15.51 & 5.21 & 6.03 & 6.21 & 8.13 & 3.24 & 11.34 & 4.60 \\ 
Qwen\text{-}audio\text{-}{chat} & 10.32 & 18.72 & 10.32 & 18.72 & 10.32 & 18.72 & 10.32 & 18.72 & 10.32 & 18.72  & 10.32 & 18.72  \\
Qwen2\text{-}audio\text{-}base & 17.99 & 8.86 & 12.64 & 17.22 & 10.85 & 24.64 & 17.36 & 13.76 & 8.95 & 17.48 & 13.56 & 16.39 \\ 
Qwen2\text{-}audio\text{-}chat & 10.64 & 18.79 & 12.11 & 19.56 & 10.76 & 18.95 & 10.41 & 19.35 & 11.60 & 19.50 & 11.10 & 19.23 \\
\bottomrule
\end{tabular}
\end{table*}

The feature extraction, which learns discriminative features via capturing audio fake artifacts from speech signals, is the key module of the pipeline detector. The features can be roughly divided into two categories \citep{DBLP:journals/corr/abs-2308-14970}: handcrafted features and deep features. Linear frequency cepstral coefficients (LFCC) is a commonly used handcrafted features that uses linear filerbanks, capturing more spectral details in the high frequency region.  \citep{DBLP:conf/icassp/YiFTNMWWTBFLWZY22,DBLP:conf/dada/YiTFYWWZZZRXZGW23}.  Nevertheless, handcrafted features are flawed by biases due to limitation of handmade representations \citep{DBLP:conf/iclr/ZeghidourTQT21}. Deep features, derived from deep neural networks, have been proposed to address these limitations. Pre-trained self-supervised speech models, such as Wav2vec2 \citep{DBLP:conf/nips/BaevskiZMA20,DBLP:conf/odyssey/TakTWJYE22} and Hubert \citep{DBLP:journals/taslp/HsuBTLSM21} are the most widely used ones \citep{DBLP:conf/odyssey/0037Y22}. \citet{DBLP:conf/odyssey/0037Y22} investigate the performance of spoof speech detection using features extracted from different self-pretrained models. The back-end classifier, tasked with learning high-level feature representations from the front-end input features, is indispensable in the audio deepfake detection. One of the widely used classifiers is LCNN \citep{DBLP:journals/tifs/WuHST18}, as it is an effective model employed as the baseline in a series of competitions, such as ASVspoof \citep{DBLP:journals/tbbis/NautschWEKVTDSY21} and ADD 2022 \citep{DBLP:conf/icassp/YiFTNMWWTBFLWZY22}.

End-to-End Models process the audio data in its raw form to capture nuanced details directly impacting audio deepfake detection performance \citep{DBLP:journals/corr/abs-2411-00121}.  Notable models include RawNet2 \citep{DBLP:conf/interspeech/JungKSKY20}, which employs Sinc-Layers \citep{DBLP:conf/slt/RavanelliB18} to extract features directly from waveforms, and RawGAT-ST, which utilizes spectral and temporal sub-graphs \citep{DBLP:journals/corr/abs-2107-12710}.  Similarly, Rawformer \citep{DBLP:conf/icassp/LiuLWLZD23} combines convolutional layers with Transformer \citep{DBLP:conf/nips/VaswaniSPUJGKP17} structures to model local and global artefacts.

\subsection{Audio Large Language Model}
In the past year, modern Large Language Models (LLMs) have demonstrated powerful reasoning and understanding abilities \citep{DBLP:journals/corr/abs-2302-13971,DBLP:journals/corr/abs-2309-16609,DBLP:journals/corr/abs-2408-13533,DBLP:journals/corr/abs-2411-18478,yu2025crisp}.  To extend the application scope of LLMs beyond pure text tasks, many LLM-based multimodal models \citep{DBLP:conf/nips/LiuLWL23a,DBLP:journals/corr/abs-2502-13923,DBLP:conf/emnlp/ZhangLZZWZQ23,DBLP:conf/iclr/0001LLKG24,DBLP:conf/emnlp/ZhangLB23,DBLP:journals/corr/abs-2502-11775} have been developed.  



For the audio modality, there have been attempts to utilize well-trained audio foundation models as tools, exemplified by AudioGPT \citep{DBLP:conf/aaai/HuangLYSCYWHHLR24} and HuggingGPT \citep{DBLP:conf/nips/0001ST00Z23}, with LLMs serving as a flexible interface.  These endeavors typically involve using LLMs to generate commands to manage external tools or to convert spoken language into text prior to LLM processing.  However, these methods often overlook critical aspects of human speech like prosody and sentiment, and struggle to handle non-verbal audio.  Such limitations pose significant challenges in effectively transferring LLM capabilities to audio applications.  Recent efforts have focused on developing end-to-end ALLMs that facilitate direct speech interaction.  SpeechGPT \citep{DBLP:conf/emnlp/ZhangLZZWZQ23} initially transforms human speech into discrete HuBERT tokens, and then establishes a three-stage training pipeline on paired speech data, speech instruction data and chain-of-modality instruction data.  LTU \citep{DBLP:conf/iclr/0001LLKG24} develops a 5M audio question answering dataset and applies supervised finetuning (SFT) to the audio modules and LoRA \citep{DBLP:conf/iclr/HuSWALWWC22} adapters of LLaMA \citep{DBLP:journals/corr/abs-2302-13971}, enhancing the model's capability to align sound perception with reasoning. Furthermore, Qwen-Audio \citep{DBLP:journals/corr/abs-2311-07919} adopts the LLaVA \citep{DBLP:conf/nips/LiuLWL23a} architecture, which has been successfully applied in vision-text LLMs, to develop a unified audio-text multi-task multilingual LLMs.

While existing ALLMs have achieved notable success in tasks such as audio caption \citep{DBLP:conf/icassp/TangYSC0LLMZ24,DBLP:journals/corr/abs-2502-15178,DBLP:journals/corr/abs-2409-12962} and speech recognition \citep{DBLP:journals/corr/abs-2408-09491,DBLP:conf/icassp/FathullahWLJSLG24,DBLP:journals/corr/abs-2311-07919}, their application in detecting fake audio remains unexplored. Therefore, in this paper, we formulate audio deepfake detection as an audio question answering task, leveraging the advanced understanding capabilities of these ALLMs to address this emerging challenge.

\section{Can ALLMs detect fake audio zero-shot?}\label{Can Audio LLMs detect?}

Recent studies \citep{DBLP:journals/corr/abs-2404-13306,DBLP:conf/cvpr/JiaLZC0JHLWL22,DBLP:journals/corr/abs-2410-09732,wang2025forensics} indicate that employing Vision Large Language Models (VLMMs) for zero-shot fake image detection presents substantial challenges, even for advanced models such as GPT4V. Despite their robust capabilities, these models often fail to achieve satisfactory performance in zero-shot fake image detection tasks. In this section, we explore whether ALLMs possess zero-shot capability for detecting fake audio.



Inspired by \citep{DBLP:conf/cvpr/JiaLZC0JHLWL22}, we utilize the following prompt templates: \textbf{Prompt1:} Is this audio fake or real? Answer fake or real.  \textbf{Prompt2:} What is the authenticity of this audio? Answer fake or real. \textbf{Prompt3:} Can you determine if this audio is fake or real?  Answer fake or real.  \textbf{Prompt4:} Tell me if this audio is a real audio?  Answer yes or no.  \textbf{Prompt5:} Please assess whether this audio recording is fake or real.  Answer fake or real. 

Our objective with these prompts is for the  ALLMs to perform binary classification, determining whether audio is "fake" or "real". This process can be described as follows: 

\begin{equation}
    res = \mathcal{M}(prompt,audio).
\end{equation}
Here, $res$ represents the response from the ALLM $\mathcal{M}$, $prompt$ and $audio$ represent the prompt template and the audio file to be assessed, respectively.


To judge the authenticity of the audio based on the model’s response $res$, we have devised a two-step process.  First, we utilize a set of rule-based standards to decide whether the audio is real or fake. For $res$ that cannot be directly classified by these rules, we conduct further assessments using \textit{gpt-3.5-turbo-0125} with the following prompt: \textit{I want to detect fake audio.  This is the answer that I get from a model: <res>.  I need you to determine whether this audio is real or fake.  If this audio is real, answer "Real".  If this audio is fake, answer "Fake".  If you can not determine, answer "Not sure".}

Based on the authenticity of the audio and the model's predictions, we categorize the audio into five classes: (1): \textbf{TP (True Positive):} The audio is real and identified as real. (2): \textbf{TN (True Negative):} The audio is fake and identified as fake. (3): \textbf{FP (False Positive):} The audio is fake but identified as real. (4): \textbf{FN (False Negative):} The audio is real but identified as fake. (5): \textbf{Fail:} The model delivers an undecidable response, exemplified by the response, "I can't determine if this audio is real or fake."


We then calculate the following evaluation metrics:
\begin{equation}
\begin{aligned}
Accuracy &= \frac{TP + TN}{\#\{\textit{total audio trials}\}}, \\
Precision &= \frac{TP}{TP + FP}, \\
mRecall &= \frac{TP}{\#\{\textit{total real trials}\}}, \\
mF_1\text{-}score &= \frac{2 \times \text{Precision} \times \text{mRecall}}{\text{Precision} + \text{mRecall}}.
\end{aligned}
\end{equation}
Here, $mRecall$ and $mF_1\text{-}score$ represent modified Recall and F1-score, respectively.


We conduct experiments on Qwen-audio series \citep{DBLP:journals/corr/abs-2311-07919} and Qwen2-audio series \citep{DBLP:journals/corr/abs-2407-10759} ALLMs. Qwen-Audio series models are multi-task ALLMs conditioning on audio and text inputs, that extends the Qwen-7B \citep{DBLP:journals/corr/abs-2309-16609} to effectively perceive audio signals by the connection of a single audio encoder. Qwen2-audio series further enhance the instruction-following capabilities by increasing the quantity and quality of data during the Supervised Fine-Tuning (SFT) stage.  Specifically, we use the following checkpoints: \textit{Qwen-audio-base\footnote{\url{https://huggingface.co/Qwen/Qwen-Audio}}}, \textit{Qwen-audio-chat\footnote{\url{https://huggingface.co/Qwen/Qwen-Audio-Chat}}}, \textit{Qwen2-audio-base\footnote{\url{https://huggingface.co/Qwen/Qwen2-Audio-7B}}}, \textit{Qwen2-audio-chat\footnote{\url{https://huggingface.co/Qwen/Qwen2-Audio-7B-Instruct}}}.

We assess the performance of ALLMs on the ASVspoof2019 LA dataset \citep{DBLP:journals/tbbis/NautschWEKVTDSY21}, a prevalent dataset in ADD research. Further details about ASVspoof2019 LA dataset are provided in Sec. \ref{dataset description}.  We report accuracy (ACC) and mF1-score with respect to different prompt templates on ASVspoof2019 LA evaluation set in Table \ref{tab:qwen-audio-performance}.  


From Table \ref{tab:qwen-audio-performance}, we observe that both the Qwen-audio series and the Qwen2-audio series ALLMs fail to effectively detect fake audio.   Specifically, the \textit{Qwen-audio-base} model exhibits an accuracy of only 11.34\% and an mF1-score of 4.60\%, averaged across five prompt templates.  Additionally, the \textit{Qwen-audio-chat} model consistently misclassifies all audio as real, regardless of the prompt used. This phenomenon underscores the challenges of relying on ALLMs for fake audio detection, stemming primarily from the fact that these ALLMs are not inherently designed for deepfake detection tasks.

\begin{figure}[t]
    \centering
    \includegraphics[width=\linewidth]{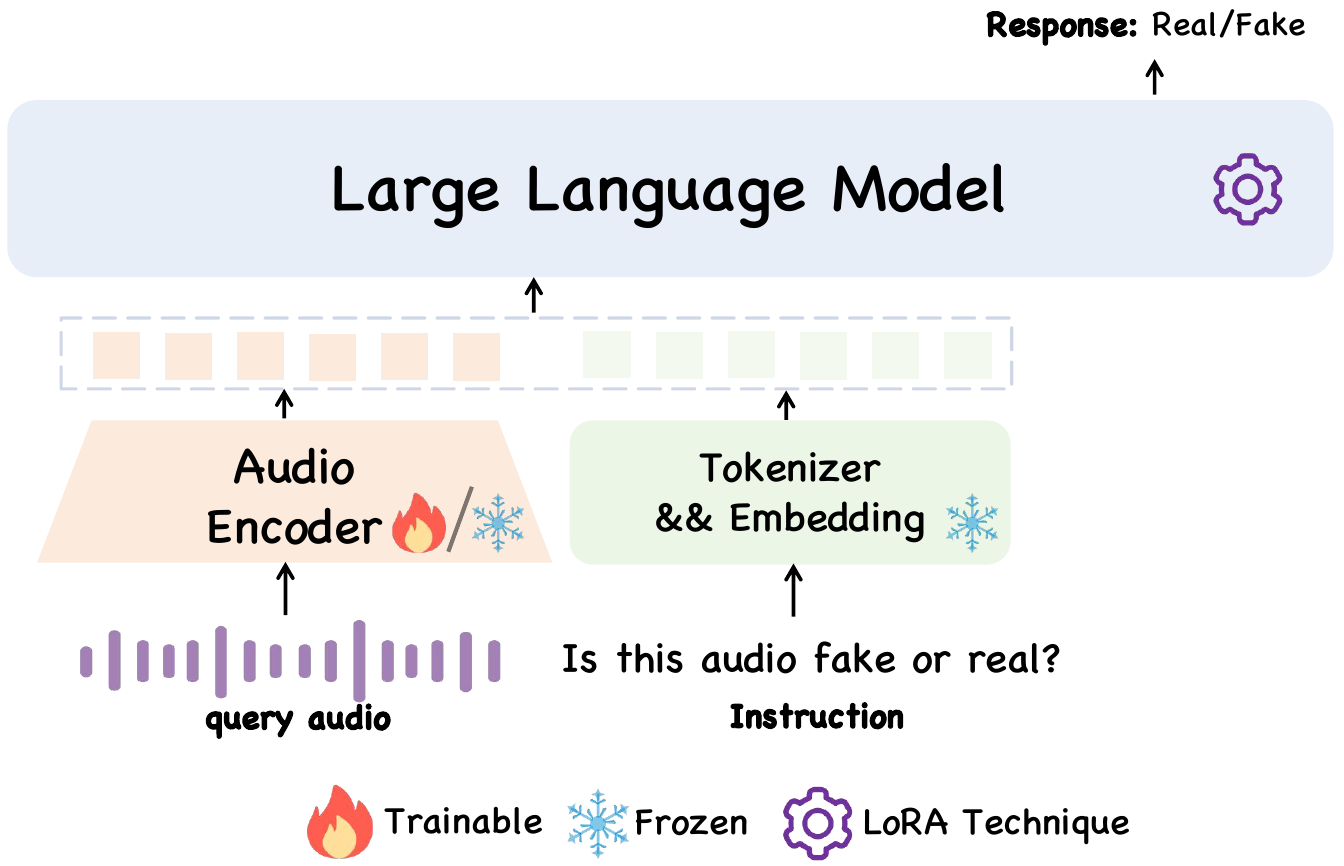} 
    \caption{Overview of our ALLM4ADD.  We reformulate audio deepfake detection as an audio question answering task.  When conducting supervised fine-tuning for audio large language model, we employ LoRA for LLM component.  For audio encoder, $\mathbf{ALLM4ADD^\triangle}$ keeps it trainable, while$\mathbf{ALLM4ADD^\star}$ keeps it frozen.}
    \label{fig:model_architecture}
\end{figure}

\section{Method}

\subsection{Task Formulation}
In order to take advantage of the Audio Large Language Model (ALLM), ALLM4ADD formulates the audio deepfake detection (ADD) task as a audio question answering (AQA) problem.  In this framework, the input comprises two crucial components: a query audio $A$ that needs to be classified as real or fake and an instruction prompt $q$, which guides ALLM4ADD in its analysis of the query audio. The instruction $q$ can take on various forms (e.g., ``Is this audio fake or real?'').  The output of this framework corresponds to the answer text $y$. While $y$ can be any text in principle, we constrain it to two options: "Fake" and "Real" during training, aligning with the ground truth to the original binary classification problem.

In summary, the ADD task can be formulated as an AQA task, which is defined as:
\begin{equation}
    \mathcal{M}(A,q) \rightarrow y,
\end{equation}
where $\mathcal{M}$ is an ALLM and the text output $y \in $ \{``Fake'', ``Real''\} corresponds to the binary result of fake audio detection.


\subsection{Model Architecture}

As shown in Figure \ref{fig:model_architecture}, ALLM4ADD adopts an architecture similar to the Qwen-Audio series, consisting of an audio encoder and a large language model. This framework takes two inputs: a query audio and a natural language instruction. Below, we describe each sub-module in detail.




\paragraph{Audio Encoder} The purpose of the audio encoder is to transform the input audio into a sequence of continuous representations. Formally, for an input sequence of raw audio signals \( A = \{ x^{(1)}, x^{(2)}, \ldots, x^{(T)} \} \), an encoder \( \mathcal{E}_{a} \) is employed to encode the audio signals $A$ into audio hidden representations $H^{a}$.  The transformation is defined as:

\begin{equation}
    H^{a} = \mathcal{E}_{a}(A), \quad H^{a} \in \mathbb{R}^{\tau \times d},
\end{equation}
 where \( \tau \) represents the output sequence length, \( d \) is the hidden size, and \( \tau \ll T \).

In practice, ALLMs predominantly employ the Whisper models \citep{DBLP:conf/icml/RadfordKXBMS23} as audio encoders. For instance, Qwen-audio uses the Whisper-large-v2 model \citep{DBLP:journals/corr/abs-2311-07919}, and Qwen2-audio opts for the Whisper-large-v3 model \citep{DBLP:journals/corr/abs-2407-10759}. These Whisper models process audio sampled at 16,000 Hz, converting it into log-Mel spectrogram representations, and have demonstrated strong performance across various speech recognition tasks.

\paragraph{Large Language Model} Our ALLM4ADD adopts a LLM as its foundation component. The model is initialized with pre-trained weights from \textit{Qwen-audio-chat} default, which is a 32-layer Transformer decoder model with a hidden size of 4096, encompassing a total of 7.7B parameters.  The LLM is employed to process audio representations and corresponding instructions, subsequently generating responses capable of discerning authenticity.  The format input to the LLM follows this format:


\begin{Verbatim}[fontsize=\small]
<|im_start|> user: <Audio> <AudioFeature> </Audio> 
Is this audio fake or real? <|im_end|> 
<|im_start|> assistant: Fake <|im_end|>
\end{Verbatim}

Here <AudioFeature> denotes audio hidden representations $H^{a}$ obtained via audio encoder $\mathcal{E}_{a}$. The special tokens <|im\_start|> and <|im\_end|> represent the beginning and the end of a sentence.

\subsection{Supervised Fine-tuning}
We construct a fine-tuning dataset $\mathcal{D}_{ft}$, comprising AQA-style, by pairing each audio with the corresponding prompt instruction.  We employ the instruction prompt $q$ as ``Is this audio fake or real?'' default.  The model's response $y$ is structured with a definitive statement "Real" if the query audio is real and "Fake" if the query audio is fake.  Consequently, $\mathcal{D}_{ft}$ is formalized as $\mathcal{D}_{ft} = \{A^i,q,y^i\}_{i=1}^{N}$.  Here $N$ represents the number of the training data. 

The initial ALLM $\mathcal{M}$ is conducted supervised fine-tuning using $\mathcal{D}_{ft}$ over the language modeling loss.  The model's objective is to minimize loss function $\mathcal{L}$ over $\mathcal{D}_{ft}$.
\begin{equation}
    \theta^* = \arg \min_{\theta} \sum_{i=1}^{N} \mathcal{L}(\mathcal{M}_\theta(A^i,q),y^i).
\end{equation}
Here $\theta$ represents the trainable parameters of ALLM $\mathcal{M}$ and $\mathcal{L}$ is the language modeling loss function.  After training on $\mathcal{D}_{ft}$, the fine-tuned ALLM $\mathcal{M}_{f}$ will be capable generating responses that could determine the authenticity of audio.

However, fine-tuning all parameters of the LLM component is time consuming and resource intensive \cite{DBLP:conf/iclr/HuSWALWWC22,DBLP:conf/coling/GuYL0Y24}. Thus, we employ the LoRA technique \cite{DBLP:conf/iclr/HuSWALWWC22}, which selectively fine-tunes a subset of the LLM's parameters, thereby forcing the model's capability on deepfake specific features while maintaining overall integrity.  Specifically, for a pre-trained weight $W_{0} \in \mathbb{R}^{m \times n}$, LoRA composes its update $\Delta{W}_{0}$ into two trainable low-rank matrices $W_{A}$ and $W_{B}$ as: $\Delta{W}_{0} =  \alpha W_{A}W_{B}$, where $W_{A} \in \mathbb{R}^{m \times r}$,$W_{B} \in \mathbb{R}^{r \times n}$,and the rank $r\ll min(m,n)$.  $W_{A}$ is initialized as a random Gaussian initialization and $W_{B}$ is initialized to all zeros at the beginning of training.  $\alpha$ serves as a hyperparameter that modulates the effect of the adaption process.  During fine-tuning, $W_{0}$ is fixed while $W_{A}$ and $W_{B}$ are trainable.  In this paper, we apply LoRA adapters to the query, key, value and output projection layers of the LLM.

Additionally, we categorize our methods based on the trainability of the audio encoder: $\mathbf{ALLM4ADD^\star}$ denotes that the encoder is frozen, while $\mathbf{ALLM4ADD^\triangle}$ indicates that the encoder is trainable.  Our motivation is to transform the feature space of the audio embeddings into one that can effectively discriminate between real and fake by training the audio encoder. This strategy aims to capture more detailed nuances of audio authenticity, thereby enhancing the model's performance in detecting fake audio.

\subsection{Evaluation}
To evaluate the performance of ALLMs in fake audio detection, we initially adopted the widely used Equal Error Rate (EER) as our primary metric \citep{DBLP:journals/corr/abs-2308-14970}. However, due to the predominance of fake samples in this task, a model might achieve a low EER by predominantly classifying samples as fake. To mitigate this issue and ensure a more comprehensive assessment, we draw inspiration from image deepfake detection methods \citep{chen2024textit,zhang2024asap} and further incorporated Accuracy (ACC) and the Area Under the Curve (AUC) as additional evaluation metrics.  Next, we describe how to compute evaluation metrics using a fine-tuned model $\mathcal{M}_{F}$.

Typically, the ALLM $\mathcal{M}_{F}$ takes as input the discrete tokens of instruction $q$ and the query audio $A$, then generates the next token $y$ as the output, which can be formulated as follows:
\begin{equation}
\begin{aligned}
    s & = \mathcal{M}_{F}(A,q) \in \mathbb{R}^V, \\
    p & = \text{Softmax}(s) \in \mathbb{R}^V, \\
\end{aligned}
\end{equation}
where $V$ is the vocabulary size, and $y$ is sampled sampled from the probability distribution $p$.

To compute our evaluation metrics, we require $\mathcal{M}_{F}$ to perform point-wise scoring for each query audio.  To this end, we conduct a bidimensional softmax over the corresponding scores of the
binary key answer words (i.e., "Fake" \& "Real"). Suppose the vocabulary indices for "Fake" and "Real" are $f$ and $r$, respectively.

Then we can obtain the probability that query audio $A$ is fake with the following formula:
\begin{equation}
    P_{r}(A \in Fake) = \frac{exp(s_{f})}{exp(s_{f}) + exp(s_{r})}.
\end{equation}

After calculating the probabilities $P_{r}(A \in Fake)$ and $P_{r}(A \in Real)$, we can compute the evaluation metrics.


\section{Experiments}

\subsection{Experiment Setup}
\subsubsection{Datasets} \label{dataset description}

The ASVspoof2019 LA dataset is a dataset for ADD, comprising 19 spoofing attack algorithms, with two types of spoofing attacks: TTS and VC.  It includes three subsets: training, development, and evaluation sets. Table \ref{statistic of dataset} presents the distribution of real and fake audio utterances across these subsets.




Inspired by the ability of multimodal large language models to quickly adapt to new tasks with limited data (\citep{DBLP:conf/nips/LiuLWL23a,DBLP:journals/corr/abs-2502-13923}), we further conduct experiments with different sampling versions of the ASVspoof2019 LA dataset. We designate the full training set as $ASV@full$, while $ASV@1/4$ represents randomly sampling one quarter of the real audio utterances and one quarter of the fake audio utterances independently from the training set. Similarly, we create $ASV@1/8$ and $ASV@1/16$ by sampling one eighth and one sixteenth of the training set, respectively. It is worth noting that we only apply this sampling process to the training set while keeping the same development and evaluation sets in all experiments to ensure fair comparison.





\begin{table}[t]
\centering
\caption{The detailed information of the ASVspoof2019 LA dataset.  The columns \# Genuine and \# Spoofed represent the number of real and fake audio utterances, respectively.}
\label{statistic of dataset}
\renewcommand{\arraystretch}{0.8}
\begin{tabular}{c|ccc}
\toprule
Set    & \# Genuine & \# Spoofed & \# Total \\ 
\midrule
Training  & 2,580      & 22,800    & 25,380  \\
Development    & 2,548      & 22,296    & 24,844  \\
Evaluation   & 7,355      & 64,578    & 71,933  \\ 
\bottomrule
\end{tabular}
\end{table}

\subsubsection{Baselines}\label{baseline description}

We compare our approach with a wide range of audio deepfake detection methods.  For conventional pipeline methods, we consider different combinations of frond-end features and back-end classifiers.  For frond-end features, we consider handcrafted features, linear frequency cepstral coefficients (LFCC) and two representative pre-trained self-supervised features: Wav2vec2.0 \citep{DBLP:conf/nips/BaevskiZMA20}, and Hubert \citep{DBLP:journals/taslp/HsuBTLSM21}. A brief introduction is provided below. \textbf{LFCC} features are derived using linear triangular filters.  We apply a 50ms window size with a 20ms shift and extract features with 60 dimensions.  \textbf{Wav2vec 2.0} \citep{DBLP:conf/nips/BaevskiZMA20} employs a convolutional encoder followed by a product quantization module to discretize audio waveform.  Then, a portion of the quantized representations is masked and modeled using a contrastive loss.   \textbf{HuBERT} \citep{DBLP:journals/taslp/HsuBTLSM21} clusters speech signals into discrete hidden units using the k-means algorithm, subsequently employing masked language modeling to predict these hidden units from masked audio segments.  For brevity, these self-supervised features are denoted as \textbf{W2V} and \textbf{Hubert}, respectively.  For back-end classifiers, we select GF \citep{DBLP:conf/odyssey/0037Y22} and LCNN \citep{DBLP:journals/tifs/WuHST18}.  \textbf{GF} \citep{DBLP:conf/odyssey/0037Y22} consists of two simple linear layers and an average pooling operation.  \textbf{LCNN} \citep{DBLP:journals/tifs/WuHST18} consists of convolutional and max-pooling layers with Max-FeatureMap (MFM) activation. 





For end-to-end models, we select five competitive methods that provide open-source code: RawNet2 \citep{DBLP:conf/icassp/Tak0TNEL21}, AASIST \citep{DBLP:journals/corr/abs-2110-01200}, RawGAT-ST \citep{DBLP:journals/corr/abs-2107-12710}, Rawformer \citep{DBLP:conf/icassp/LiuLWLZD23} and RawBMamba \citep{DBLP:journals/corr/abs-2406-06086}.  \textbf{RawNet2} \citep{DBLP:conf/icassp/Tak0TNEL21} is a convolutional neural network operating directly on raw audio waveforms, utilizing residual blocks and Sinc-Layers \citep{DBLP:conf/slt/RavanelliB18} as band-pass filters for effective ADD.  \textbf{AASIST} \citep{DBLP:journals/corr/abs-2110-01200} employs a heterogeneous stacking graph attention layer to model artifacts across temporal and spectral segments.  \textbf{RawGAT-ST} \citep{DBLP:journals/corr/abs-2107-12710} utilizes spectral and temporal sub-graphs integrated with a graph pooling strategy, effectively processing complex auditory environments.  \textbf{Rawformer} \citep{DBLP:conf/icassp/LiuLWLZD23} integrates convolution layer and transformer to model local and global artefacts and relationship directly on raw audio.  \textbf{RawBMamba} \citep{DBLP:journals/corr/abs-2406-06086} proposes an end-to-end bidirectional state space model to capture both short- and long-range discriminative information.



\subsubsection{Evaluation Metric}
To comprehensively assess the effectiveness of our method, following \citep{chen2024textit,zhang2024asap}, we select Equal Error Rate (EER), Accuracy (ACC), and Area Under the Curve (AUC) as evaluation metrics for fake audio detection. \textbf{Lower EER values and higher ACC and AUC scores indicate better fake audio detection performance}.


\begin{table*}[htbp]
\centering
\caption{Performance comparison of our methods with conventional pipeline and end-to-end models.  (Train) and (Frozen) represent whether the self-supervised features are trainable.  $\mathbf{ALLM4ADD^\star}$ and $\mathbf{ALLM4ADD^\triangle}$ denote the audio encoder is frozen and trainable, respectively.  We train the models across: ASV@full, ASV@1/4, ASV@1/8, and ASV@1/16, and report EER (\% ), ACC (\% ), and AUC (\% ) on the ASVspoof2019 LA evaluation set. Best results in each column are highlighted in bold.}
\label{tab:model-performance}
\renewcommand{\arraystretch}{0.82}
\begin{tabular}{c|cccccccccccc}
\toprule
\multirow{2}{*}{Models} & \multicolumn{3}{c}{$ASV@full$} & \multicolumn{3}{c}{$ASV@1/4$} & \multicolumn{3}{c}{$ASV@1/8$} & \multicolumn{3}{c}{$ASV@1/16$}\\
\cmidrule(r){2-4} \cmidrule(r){5-7} \cmidrule(r){8-10} \cmidrule(r){11-13}
 & EER \lz{$\downarrow$} & ACC \lz{$\uparrow$} & AUC \lz{$\uparrow$} & EER \lz{$\downarrow$} & ACC \lz{$\uparrow$} & AUC \lz{$\uparrow$} & EER \lz{$\downarrow$} & ACC \lz{$\uparrow$} & AUC \lz{$\uparrow$} & EER \lz{$\downarrow$} & ACC \lz{$\uparrow$} & AUC \lz{$\uparrow$}\\
\midrule
\multicolumn{13}{c}{\textit{End-to-End Methods}} \\
\midrule
Rawnet2 & 4.32 & 94.07 & 99.07 & 7.92 & 93.26 & 97.26 & 9.84 & 92.12 & 96.01 & 15.05 & 90.82 & 92.27 \\
AASIST &  0.83 &  95.18 & 99.92 & 2.93 & 96.83 & 99.49 & 3.96 & 96.25 & 99.32 & 5.54 & 94.13 & 98.68 \\
RawGAT-ST & 1.71 & 97.35 & 99.55 & 3.01 & 96.69 & 99.17 & 4.43 & 93.26 & 98.45 & 10.49 & 95.08 & 96.40 \\
Rawformer & 1.07 & 99.01 & 99.79 & 2.27 & 98.42 & 99.56 & 4.09 & 96.92 & 99.18 & 5.42 & 96.98 & 98.70 \\
RawBMamba & 1.19 & 98.67 & 99.87 & 5.21 & 93.03 & 98.72 & 7.28 & 92.21 & 97.69 & 9.54 & 91.36 & 96.39 \\
\midrule  
\multicolumn{13}{c}{\textit{Conventional Pipeline Methods}} \\
\midrule
W2V+GF (Frozen) & 6.23 & 94.84 & 98.55 & 8.64 & 92.71 & 97.33 & 8.98 & 91.21 & 97.13 & 10.14 & 91.99 & 96.23 \\
W2V+GF (Train) & 2.09 & 96.51 & 99.77 & 3.85 & 96.04 & 99.41 & 4.08 & 95.55 & 99.31 & 4.73 & 96.68 & 98.82 \\
HuBert+GF (Frozen) &  8.21 & 93.54 & 97.67 & 8.82 & 90.93 & 97.28 & 9.74 & 92.30 & 96.69  & 11.28 & 89.74 & 95.81 \\
HuBert+GF (Train) &  1.94 & 95.22 & 99.58 & 3.23 & 95.48 & 99.41 & 4.28 & 95.51 & 99.05 & 5.60 & 95.39 & 98.16 \\
LFCC+LCNN & 3.89 & 95.59 & 99.14 & 5.74 & 94.74 & 98.51 & 7.90 & 91.50 & 98.18 & 10.86 & 90.91 & 95.70 \\
W2V+LCNN (Frozen) & 3.28 & 97.43 & 99.52 & 5.11 & 96.59 & 98.83 & 7.14 & 95.63 & 98.12 & 9.65 & 92.29 & 96.50 \\
\midrule
\multicolumn{13}{c}{\textit{ALLM-based Methods}} \\
\midrule
$ALLM4ADD^\star$ & 1.30 & 99.26 & 99.88 & 1.97 & 98.87 & 99.65 & 2.46 & $\mathbf{98.69}$ & 99.54 & 3.13 & 98.11 & $\mathbf{99.43}$ \\
$ALLM4ADD^\triangle$ & $\mathbf{0.41}$ & $\mathbf{99.39}$ & $\mathbf{99.97}$ & $\mathbf{1.15}$ & $\mathbf{99.12}$ & $\mathbf{99.71}$ & $\mathbf{1.63}$ & 98.55 & $\mathbf{99.75}$ & $\mathbf{2.45}$ & $\mathbf{98.52}$ & 99.39 \\
\bottomrule
\end{tabular}
\end{table*}

\subsection{Implementation Details}
We use \textit{Qwen-audio-chat} weights as our default initial weights.  We optimize our model using the Adam optimizer with hyperparameters \( \beta = (0.9, 0.95) \) and implement a cosine learning rate scheduler. The warm-up ratio is set at 0.01 for $ASV@full$, $ASV@1/4$, and $ASV@1/8$ configurations, and 0.05 for $ASV@1/16$.   We investigate two versions of our ALLM4ADD: $ALLM4ADD^\star$ and $ALLM4ADD^\triangle$. For  $ALLM4ADD^\star$, the audio encoder is frozen; for $ALLM4ADD^\triangle$, the audio encoder is trainable. The initial learning rate is determined through beam search within the range of [3e-5, 4e-5, 5e-5, 1e-4].  Additionally, we apply a weight decay of 0.1 and a gradient clipping threshold of 1.0 to maintain training stability. Training epochs varies: 5 epochs for $ASV@full$, $ASV@1/4$, and $ASV@1/8$, and increase to 10 epochs for $ASV@1/16$.  Furthermore, our training processes utilize bfloat16 precision with automatic mixed precision for efficiency.  We conduct all experiments on a single Nvidia A100.  For LoRA configuration, we configure our LoRA hyperparameter as follows: LoRA rank $r$ as 64, LoRA alpha (scaling factor) as 16 and LoRA dropout as 0.05. We apply LoRA to the query, key, value, and output projection layers.




For conventional pipeline baselines, we adhere to the hyperparameter provided in \citep{DBLP:conf/odyssey/0037Y22}.  For end-to-end baselines, we adhere to the official codebase and train the models for 100 epochs.  To maintain a fair comparison across all methods, we do not employ data augmentation techniques.



\subsection{Experimental Results}\label{main Experimental Results}
To demonstrate the superiority of our ALLM-based fake audio detection method, we compare it with extensive baselines detailed in Sec. \ref{baseline description}. The models are trained on $ASV@full$, $ASV@1/4$, $ASV@1/8$, and $ASV@1/16$, and are evaluated on the ASVspoof2019 evaluation set.  From Table \ref{tab:model-performance}, we can draw the following observations:


\begin{itemize}[leftmargin=*]
    \item  Our ALLM4ADD can achieve excellent results in ADD task. When trained on $ASV@full$, $ALLM4ADD^\triangle$ achieves an EER of 0.41\%, surpassing the best performances of end-to-end and conventional pipeline baselines, which are 0.83\% and 1.94\%, respectively.  Furthermore, ALLM4ADD maintains strong performance when data is scarce. Specifically, under the $ASV@1/16$ setting, $ALLM4ADD^\triangle$ still achieves an EER of 2.45\% and an accuracy of 98.52\%.  These results demonstrate the superiority of our ALLM4ADD for the ADD task.
    \item  As the training data decreases from $ASV@full$ to $ASV@1/16$, the performance drop of ALLM4ADD is more moderate compared to that observed with end-to-end and conventional pipeline baselines. For example, $ALLM4ADD^\star$ observes an EER increase from 1.30\% to 3.13\%, and $ALLM4ADD^\triangle$ from 0.41\% to 2.45\%, whereas the AASIST model suffers a substantial increase in EER from 0.83\% to 5.54\%.  We attribute this phenomenon to the ALLM's ability to learn and apply transferable skills from limited data.
    \item Training the audio encoder can often lead to improved performance. When averaged across $ASV@full$, $ASV@1/4$, $ASV@1/8$, and $ASV@1/16$ settings, the EER of $ALLM4ADD^\triangle$ is 0.81\% lower than that of $ALLM4ADD^\star$.  We speculate that this improvement stems from the encoder's trainability, allowing it to extract more information relevant to audio authenticity.    
    \item We observe that some models, such as the AASIST model trained under $ASV@full$, exhibit a low EER value at 0.83\%, yet its accuracy remains relatively low at only 95.18\%.  We suggest that research in fake audio detection should consider multiple evaluation metrics to ensure a comprehensive evaluation.
\end{itemize}

\begin{figure*}[htbp]  
\centering  
\includegraphics[width=0.7\textwidth]{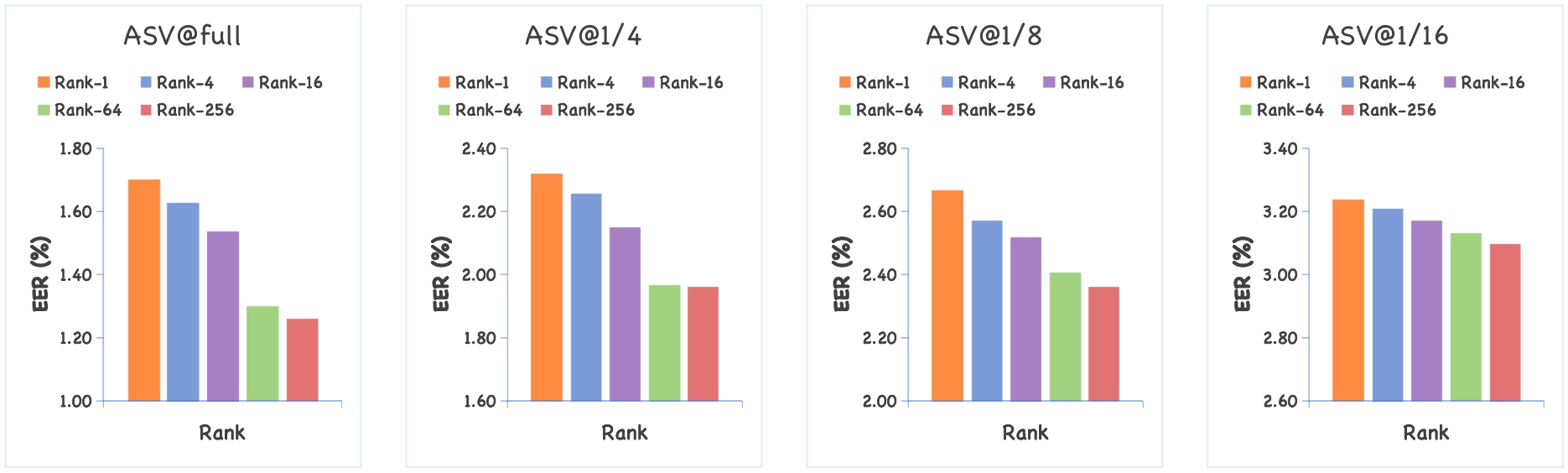}
\caption{We conduct experiments using ranks \{1, 4, 16, 64, 256\}, and the corresponding EER (\%) on ASVspoof2019 LA evaluation set under ASV@full, ASV@1/4, ASV@1/8, and ASV@1/16 settings are depicted.}  
\label{fig:rank experment}  
\end{figure*}

\section{Ablation Study}
This section investigates the following research questions (Qs).
\begin{itemize}[leftmargin=*]
    \item \textbf{Q1:} How is the model's generalization performance?
    \item \textbf{Q2:} What is the impact of different prompt templates?
    \item \textbf{Q3:} What is the impact of different LoRA ranks?
     \item \textbf{Q4:} What is the impact of different ALLM backbones?
    \item \textbf{Q5:} How is the model's performance on other fake type datasets?
    \item \textbf{Q6:} How is the model's performance on extremely limited data?
\end{itemize}

\subsection{Generalization Capabilities of Models (Q1)}
\subsubsection{Experimental Setup.}
To assess the ability of our model to generalize to real-world fake audio samples, we evaluate its performance on the In-the-Wild dataset \citep{DBLP:conf/interspeech/MullerCDFB22} containing 19,963 genuine audio files and 11,816 fake audio files. Specifically, we evaluate the performance of our models and several baseline models trained under $ASV@full$ and $ASV@1/16$ settings on the In-the-Wild dataset. The experimental results are presented in Table \ref{generation capability of llm.}.

\subsubsection{Experimental Results.}

Table \ref{generation capability of llm.} illustrates that $ALLM4ADD^\triangle$ outperforms both the end-to-end and conventional pipeline baselines on the In-the-Wild dataset. Specifically, when trained on $ASV@full$, $ALLM4ADD^\triangle$  demonstrates an EER of 26.99\%, whereas the best results from the end-to-end and pipeline models are 36.11\% and 30.54\%, respectively.

\begin{table}[t]
\centering
\caption{Comparison of our models with baselines on the In-the-Wild dataset, reporting EER (\%), ACC (\%), and AUC (\%). The Best results in each column are highlighted in bold.}
\label{generation capability of llm.}
\setlength{\tabcolsep}{1.8pt}
\renewcommand{\arraystretch}{0.7}
\begin{tabular}{c|ccc|ccc}
\toprule
 \multirow{2}{*}{Models} & \multicolumn{3}{c|}{$ASV@full$} & \multicolumn{3}{c}{$ASV@1/16$} \\
\cmidrule(r){2-7}
 & EER \lz{$\downarrow$} & ACC \lz{$\uparrow$} & AUC \lz{$\uparrow$} & EER \lz{$\downarrow$} & ACC \lz{$\uparrow$} & AUC \lz{$\uparrow$}\\
\midrule
$ALLM4ADD^\star$ & 32.04 & 65.28 & 73.82 & 45.12 & 40.66 & 51.64 \\ 
$ALLM4ADD^\triangle$ & $\mathbf{26.99}$ & $\mathbf{80.28}$ & $\mathbf{90.83}$ & $\mathbf{33.20}$ & $\mathbf{51.84}$ & 57.89 \\ 
\midrule
AASIST & 43.02 & 55.98 & 59.18 & 44.50 & 40.47 &  57.11 \\ 
Rawformer & 49.22 & 44.63 & 51.78 & 52.46 & 47.08 & 45.48\\ 
\midrule
LFCC + LCNN & 78.42 & 25.21 & 14.11 & 75.05 & 30.01 & 17.33  \\ 
Hubert + GF (Train) & 30.54 & 52.53 & 76.80 & 39.93 & 39.93 & $\mathbf{67.17}$ \\
\bottomrule
\end{tabular}
\end{table}

\begin{table}[t]
\centering
\caption{EER (\%) of different prompt templates across ASV@full, ASV@1/4, ASV@1/8, and ASV@1/16.  The best results in each column are highlighted in bold.}
\label{tab:prompt_table}
\setlength{\tabcolsep}{3pt}
\renewcommand{\arraystretch}{0.7}
\begin{tabular}{ccccc}
\toprule
\textbf{Template} & $ASV@full$ & $ASV@1/4$ & $ASV@1/8$ & $ASV@1/16$ \\ 
\midrule
prompt1 & \textbf{1.30} & \textbf{1.97} & \textbf{2.46} & 3.13 \\ 
prompt2 & 1.35 &  2.05 & 2.60 & 3.09 \\ 
prompt3 & 1.39 &  2.15 & 2.49 & 3.16 \\
prompt4 & 1.38 &  2.02 & 2.51 & \textbf{3.07}\\ 
prompt5 & 1.36 &  2.08 & 2.50 & 3.14 \\ 
\bottomrule
\end{tabular}
\end{table}

\subsection{Effect of Prompt Templates (Q2)} \label{Effect of Different Prompt Templates}

\subsubsection{Experimental Setup.}
In this section, we aim to investigate the impact of different prompt templates on ADD performance.  We employ prompts 1 to 5 as described in Sec. \ref{Can Audio LLMs detect?}, omitting the final sentence "Answer fake or real.".  To ensure that the results predominantly reflect the interaction between the prompt templates and the LLM, we froze the audio encoder and solely fine-tune the LLM component using the LoRA technique, which corresponds to $ALLM4ADD^\star$.  EER (\%) on ASVspoof2019 LA evaluation set across different settings are presented in Table \ref{tab:prompt_table}.

\subsubsection{Experimental Results.}
Experimental results shown in Table \ref{tab:prompt_table} reveal that although there are slight variations in performance between the different prompt templates, all templates consistently achieve satisfactory results.  Furthermore, we find that, except under the $ASV@1/16$ setting, prompt1 consistently yields the best results. 



\subsection{Effect of LoRA Rank (Q3)}
\subsubsection{Experimental Setup.}
To evaluate the impact of different LoRA ranks, we explore the rank with values $\{1, 4, 16, 64, 256\}$. Following Sec. \ref{Effect of Different Prompt Templates}, we employ $ALLM4ADD^\star$ for experiments.  Performance is evaluated using the ASVspoof2019 LA evaluation set, with EER (\%) measured across the $ASV@full$, $ASV@1/4$, $ASV@1/8$, and $ASV@1/16$ settings. Experimental results are depicted in Figure \ref{fig:rank experment}.


\subsubsection{Experimental Results.}
The results presented in Table 1 demonstrate a progressive decrease in EER with increasing rank, indicating that the incorporation of more trainable parameters allows our method to discern finer details in fake audio, thus enhancing the effectiveness of the audio deepfake detection system.  Additionally, the impact of increasing rank on model performance is more pronounced when trained on $ASV@full$ setting. 



\subsection{Effect of ALLM backbones (Q4)}

\subsubsection{Experimental Setup.}
To explore the impact of different ALLM backbones, we compare the performance of the \textit{Qwen-audio-chat} and \textit{Qwen-audio-base} backbone at $ASV@full$ and $ASV@1/16$ settings. Table \ref{tab:performance on diffrenet audio backbone} presents the performance on ASVspoof2019 LA evaluation set. $ALLM4ADD^\star$ indicates that the audio encoder is frozen while $ALLM4ADD^\triangle$ denotes that it is trainable. Further evaluation of more ALLMs is reserved for future work.

\begin{table}[t]
\centering
\setlength{\tabcolsep}{2pt}
\caption{Comparison of different ALLM backbones.  Base and Chat denote Qwen-audio-base and Qwen-audio-chat backbones,  respectively.}
\label{tab:performance on diffrenet audio backbone}
\renewcommand{\arraystretch}{0.8}
\begin{tabular}{c|ccc|ccc}
\toprule
 \multirow{2}{*}{Methods}  & \multicolumn{3}{c|}{$ASV@full$} & \multicolumn{3}{c}{$ASV@1/16$} \\
 & EER \lz{$\downarrow$} & ACC \lz{$\uparrow$} & AUC \lz{$\uparrow$} & EER \lz{$\downarrow$} & ACC \lz{$\uparrow$} & AUC \lz{$\uparrow$} \\
\midrule
$ALLM4ADD^\star$(Base) & 1.78 & 98.33 & 99.65 & 4.19 & 97.84 & 99.10 \\
$ALLM4ADD^\star$(Chat) & 1.30 & 99.26 & 99.88 & 3.13 & 98.11 & 99.43 \\
\midrule
$ALLM4ADD^\triangle$(Base) & 1.29 & 98.25 & 99.71 & 3.79 & 98.63 & 99.21 \\
$ALLM4ADD^\triangle$(Chat) & 0.41 & 99.39 & 99.97 & 2.45 & 98.52 & 99.21 \\
\bottomrule
\end{tabular}
\end{table}

\subsubsection{Experimental Results.}
From Table \ref{tab:performance on diffrenet audio backbone}, we observe that the \textit{Qwen-audio-chat} backbone consistently outperforms the \textit{Qwen-audio-base} backbone under both the $ASV@full$ and $ASV@1/16$ settings,  regardless of whether the audio encoder is trainable.  For example, under $ALLM4ADD^\triangle$ setting, the \textit{Qwen-audio-chat} backbone achieves an EER of 0.41\%, compared with 1.29\% for the \textit{Qwen-audio-base} backbone when trained on $ASV@full$.

\begin{table}[t]
\centering
\caption{Comparison of our method and baselines on the SceneFake and EmoFake datasets.  We report EER (\%) for both @full and @1/16 settings on SceneFake and EmoFake.}
\label{emofake and scenefake}
\renewcommand{\arraystretch}{0.7}
\begin{tabular}{c|cc|cc}
\toprule
\multirow{2}{*}{Dataset} & \multicolumn{2}{c|}{Scenefake} & \multicolumn{2}{c}{Emofake} \\ 
\cmidrule(r){2-3} \cmidrule(r){4-5}
 & $@full$ & $@1/16$ & $@full$ & $@1/16$ \\ 
\midrule
$ALLM4ADD^\star$ & 9.10 & 14.53 & 1.00 & 1.33 \\ 
$ALLM4ADD^\triangle$ & $\mathbf{8.14}$ & $\mathbf{11.05}$ & $\mathbf{0.86}$ & $\mathbf{1.28}$ \\ 
\midrule
AASIST & 13.43 & 17.97 & 1.22 & 2.82 \\ 
Rawformer & 11.38 & 16.31 & 2.68 & 3.58 \\ 
\midrule
LFCC + LCNN & 12.72 & 16.80 &  5.86 &  11.05 \\ 
Hubert+GF (Train) & 12.53 & 17.96 & 3.66 & 10.07 \\
\bottomrule
\end{tabular}
\end{table}

\subsection{Performance on other Fake Datasets (Q5)}
\subsubsection{Experimental Setup.}
In this section, we aim to explore the performance of ALLM4ADD  on different types of ADD datasets. Specifically, we focus on the EmoFake \citep{DBLP:conf/cncl/ZhaoYTWD24} and SceneFake \citep{DBLP:journals/pr/YiWTZFTMF24} datasets. EmoFake involves modifying the emotional characteristics of speech while preserving other information. SceneFake, on the other hand, entails altering the acoustic scene of an utterance via speech enhancement techniques, without changing other aspects.  We evaluate both $ALLM4ADD^\star$ and $ALLM4ADD^\triangle$, alongside several baselines under both $@full$ and $@1/16$ settings.  For the $@1/16$ setting of EmoFake and SceneFake, we extract 1/16 of the real and fake audio files from the corresponding training set.  Performance on the corresponding evaluation set is presented in Table \ref{emofake and scenefake}.
\subsubsection{Experimental Results.}
We observe that the $ALLM4ADD^\triangle$ consistently achieves the lowest EER across both SceneFake and EmoFake datasets, under both $@full$ and $@1/16$ settings.  Specifically, at the $@full$ setting, $ALLM4ADD^\triangle$ exhibits an EER of 8.14\% for SceneFake and 0.86\% for EmoFake, compared to the best EERs of competing end-to-end models, which are 11.38\% and 1.22\%, respectively.  This consistent superior performance underscores the effectiveness of our approach, validating its capability in detecting various types of audio deepfakes.

\begin{figure}[t]
    \centering
    \includegraphics[width=0.65\linewidth]{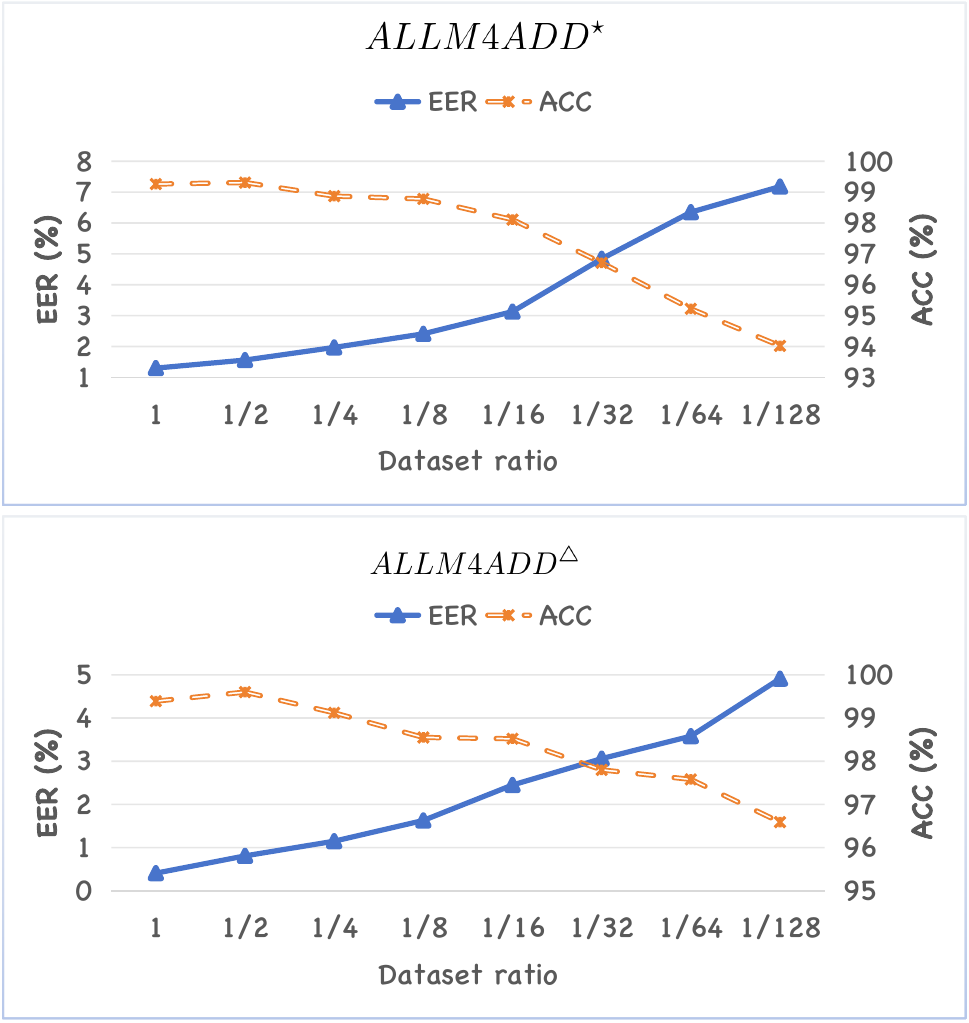} 
    \caption{Performance on ASVspoof 2019 LA evaluation set across different data ratios.  We depict EER (\%) and AUC (\%) of $\mathbf{ALLM4ADD^\triangle}$ and $\mathbf{ALLM4ADD^\star}$.}
    \label{fig:ratio of methods}
\end{figure}

\subsection{Performance on extremely limited data (Q6)}

\subsubsection{Experimental Setup.}
Although we report the performance of ALLM4ADD under various training set sizes in Sec. \ref{main Experimental Results}, the model's effectiveness on extremely limited data volume is not explored. To address this, we conduct further experiments with training sets sized at fractions \{1, 1/2, 1/4, 1/8, 1/16, 1/32, 1/64, 1/128\} of the total data volume in this section.  For $ASV@1/32$, $ASV@1/64$, and $ASV@1/128$ settings, we train the models for 15 epochs and adjust the learning rate according to the data volume to ensure the best performance.  Notably, under the $ASV@1/128$ setting, our training dataset consists of only 178 fake audio samples and 20 real audio samples.  Figure \ref{fig:ratio of methods} illustrates the EER (\%) and ACC (\%) of $ALLM4ADD^\triangle$ and $ALLM4ADD^\star$ across these proportions.


\subsubsection{Experimental Results.}

From Figure \ref{fig:ratio of methods}, we draw the following observations: (1) The performance of the model tends to decline as the data ratio decreases. (2) Our method still achieves impressive performance even in settings with extremely scarce data. For instance, in the $ASV@1/128$ setting, \textbf{$ALLM4ADD^\triangle$ still maintains an EER below 5\% and an accuracy over 96\%}. This can likely be attributed to the ALLM's robust few-shot capabilities, which enable it to adapt effectively to downstream tasks with little additional training. These experimental results demonstrate the feasibility of effective audio deepfake detection in scenarios with minimal data.

\section{Conclusion}



This paper presents our pioneering work on applying ALLMs to ADD.  First, we conduct a comprehensive evaluation of ALLMs' zero-shot capabilities for fake audio detection, revealing their limitations on the ADD task.  We then propose ALLM4ADD, a novel framework that reformulates the ADD task as an AQA problem, to endow ALLMs with the ability to detect fake audio.  Extensive empirical results demonstrate that ALLM4ADD can achieve superior performance compared to existing methods, particularly in data-scarce scenarios.  Notably, our method can achieve an EER below 5\% and an accuracy over 96\% with only around 200 training samples.  These findings underscore the potential of ALLM-based approaches in advancing fake audio detection.  In the future, we plan to leverage ALLMs to develop an ADD system that unifies deepfake detection with explaining capabilities.

\section{Acknowledgements}
This work is supported by the National Natural Science Foundation of China (NSFC) (No. 62322120, No.U21B2010, No. 62306316, No. 62206278).



\bibliographystyle{ACM-Reference-Format}
\bibliography{sample-base}

\appendix
\end{document}